\documentclass[11pt]{article}
\usepackage{hyperref, graphics}
\pdfoutput=1
\begin{document}
\title{Dead Waters: Large amplitude interfacial waves generated by a boat in a stratified fluid}
\author{Romain Vasseur, Matthieu Mercier, Thierry Dauxois \\
\\\vspace{6pt} Universit\'e de Lyon, Laboratoire de Physique,
ENS Lyon, CNRS,\\ 46 all\'{e}e d'Italie, 69364 Lyon cedex 07,
France} \maketitle
\begin{abstract}
We present fluid dynamics videos of the motion of a boat on a
two-layer or three-layer fluid. Under certain specific conditions,
this setup generates large amplitude interfacial waves, while no
surface waves are visible. The boat is slowed down leading to a
peristaltic effect and sometimes even stopped: this is the so-called
dead water phenomenon.
\end{abstract}

The phenomenon of dead water was first encountered by F. Nansen on
the Barentsz Sea, and later carefully studied by Ekman~\cite{Ekman}.
It has been recently studied in the two layer situation by Leo
Maas~\cite{Maas}.

We performed experiments~\cite{RomainMatthieu} by studying a toy
boat moving in a long and thin plexiglass tank
$300\,\textrm{x}\,41\,\textrm{x}\,10.4\,\textrm{cm}^3$, filled with
several layers of water with different densities. The boat is pulled
by a thin rope (invisible in the video) at constant horizontal
force. Its motion is recorded with a video camera approximately 4 m
to the side of the tank. We present first a setup with two different
layers. The 5 cm upper fresh water layer, colored in red, has a
density $\rho_1$ = 1.000(7) g.cm$^{-3}$, while the 12 cm transparent
bottom salted one corresponds to  $\rho_2$ = 1.022(5) g.cm$^{-3}$.

At the beginning of the
\href{http://ecommons.library.cornell.edu/bitstream/1813/11470/3/VasseurMercierDauxois_mpeg1_APS2008.mpg}{Video
1} or
\href{http://ecommons.library.cornell.edu/bitstream/1813/11470/2/VasseurMercierDauxois_mpeg2_APS2008.mpg}{Video
2}, the boat reaches a constant velocity almost immediately, while
one distinguishes, at the rear of the boat, no surface wave but a
striking interfacial wave. The wave, generated by the motion of the
boat, increases its amplitude and therefore its speed. Consequently,
the wave catches up to the boat and slows it down. Sometimes it is
even stopped. The wave breaks because it hits the boat. The boat
being released from its trap, it starts its motion again, and the
process repeats.

\begin{figure}[htb]
\begin{center}
\resizebox{0.8\textwidth}{!}{\includegraphics{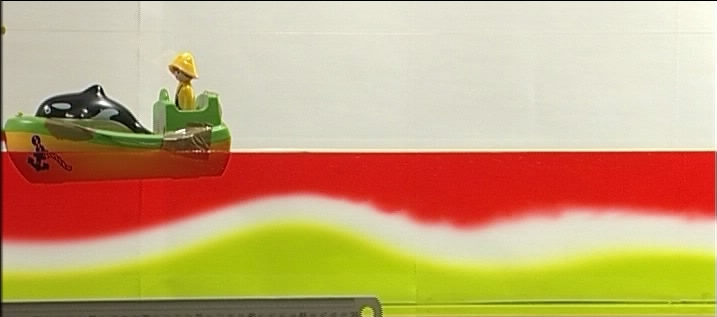}}
\end{center}
\caption{(Color) Large amplitude interfacial waves generated by a
boat moving on a three-layer stratified fluid. Note the total
absence of surface waves.} \label{beteeth}
\end{figure}

The second part of the video shows that the phenomenon persists in
the three-layer situation. The interfacial wave generated in the
upper interface is transferred to the bottom one (see
Fig.~\ref{beteeth}). One can see the boat might be stopped again.
The spatio-temporal plot which emphasizes the different waves
emitted in front or at the back of the boat, shows that a nice
quantitative analysis can thus be performed. Several patterns of the
waves have been identified: i) in-phase and ii) out-of-phase
interfacial waves, iii)~soliton-like excitation emitted in front of
the boat when it stops.

\bigskip{\bf Acknowledgments}

We warmly thank Leo Maas for suggestions and Playmobil$\copyright$
for the quality of the boat. This work has been partially supported
by 2005-ANR project TOPOGI-3D.


\begin{thebibliography}{10}
\bibitem{Ekman}
V.W. Ekman. {\em On dead water}. Norw. N. Polar Exped. 1893-1896:
Sci Results, XV, Christiana, (1904).

\bibitem{Maas}
L. Maas, NIOZ, Netherland Institute of Sea Research, unpublished
(2005).

\bibitem{RomainMatthieu}
R. Vasseur, M. Mercier, T. Dauxois, ``Interfacial Waves in
stratified fluids", to be published (2008).


\end{thebibliography}
\end{document}